\documentclass[pdftex,twocolumn,epjc3]{svjour3}          

\RequirePackage[T1]{fontenc}

\smartqed  

\RequirePackage{graphicx}
\RequirePackage{mathptmx}      
\RequirePackage{flushend}
\RequirePackage[numbers,sort&compress]{natbib}
\RequirePackage[colorlinks,citecolor=blue,urlcolor=blue,linkcolor=blue]{hyperref}

\journalname{Eur. Phys. J. C}

\usepackage{textcase}
\usepackage{amssymb,amsmath,mathtools} 
\usepackage{url}
\usepackage{natbib}

\usepackage{dcolumn}
\usepackage{bm}
\usepackage[T1]{fontenc}
\usepackage[utf8]{inputenc}
\usepackage{booktabs}
\usepackage{multirow}
\usepackage{graphicx}
\usepackage{xcolor}
\usepackage{microtype} 
\usepackage{hyperref}

\hypersetup{
  colorlinks=true,
  linkcolor=red,
  citecolor=blue,
  urlcolor=blue
}

\begin{document}
\title{Extragalactic test of General Relativity with time-delay gravitational lenses}

\author{Wuzheng Guo\thanksref{addr1,addr2}
        \and
        Shuo Cao$\ast$\thanksref{e1,addr1,addr2} 
        \and
        Qiumin Wang$\dagger$\thanksref{e2,addr1,addr2}
        \and
        Yun Chen\thanksref{addr3}
        \and
        Marek Biesiada$\ddagger$\thanksref{e3,addr4}
        \and
        Tonghua Liu\thanksref{addr5}
        \and
        Yujie Lian\thanksref{addr1,addr2}
}

\thankstext{e1}{e-mail: caoshuo@bnu.edu.cn}
\thankstext{e2}{e-mail: wangqiumin@mail.bnu.edu.cn}
\thankstext{e3}{e-mail: Marek.Biesiada@ncbj.gov.pl}

\institute{Institute for Frontiers in Astronomy and Astrophysics, Beijing Normal University, Beijing 102206, China\label{addr1}
          \and
          School of Physics and Astronomy, Beijing Normal University, Beijing 100875, China \label{addr2}
          \and
          Key Laboratory for Computational Astrophysics, National Astronomical Observatories, Chinese Academy of Sciences, Beijing 100101, China \label{addr3}
          \and
          National Centre for Nuclear Research, Pasteura 7, PL-02-093 Warsaw, Poland\label{addr4}
          \and
          School of Physics and Optoelectronic, Yangtze University, Jingzhou 434023, China\label{addr5}
}
\date{Accepted: May 15,2026}

\maketitle

\begin{abstract}
Strong gravitational lensing, a key prediction of General Relativity (GR), offers a unique environment for examining alternative modified gravity theories. In this Letter, we employ a model-independent approach to estimate the parameterized post-Newtonian parameter $\gamma_{\rm PPN}$ using the time-delay measurements from H0LiCOW strong lensing systems. To minimize potential biases from cosmological models in testing GR, we use Gaussian Process regression (GPR) to reconstruct angular diameter distances ($D_{\rm A}$) from the newest baryon acoustic oscillation (BAO) measurements, provided by the Dark Energy Spectroscopic Instrument (DESI) DR2 data. Based on the reconstructed angular diameter distances and four H0LiCOW lenses, we directly estimate the post-Newtonian parameter $\gamma_{\rm PPN}=0.93^{+0.16}_{-0.17}$ and the sound horizon scale $r_{\rm d}=136.36^{+5.14}_{-3.20}~{\rm Mpc}$. This is the first simultaneous measurement of $\gamma_{\rm PPN}$ and $r_{\rm d}$ without any assumptions about the contents of the universe or the theory of gravity. In the new framework of distance ratio $D_{\Delta t}/D_{\rm l}$ which avoids the bias introduced by $r_{\rm d}$, the $\gamma_{\rm PPN}$ constraint can be further improved to $\gamma_{\rm PPN}=0.89^{+0.19}_{-0.15}$. Our results provide a direct test of GR at the extragalactic scale, which is well consistent with the prediction of GR within $1\sigma$.

\end{abstract}

\maketitle


\section{Introduction}

Einstein's General Relativity (GR), which stands as the cornerstone of modern cosmology, has laid the foundation for the Big Bang theory and the expanding universe model, shaping our comprehension of cosmic evolution. As a successful implementation of GR, the Lambda Cold Dark Matter ($\Lambda$CDM) model aligns well with various astronomical observations, including Type Ia supernovae (SN Ia) \citep{Riess2007} and the cosmic microwave background (CMB) \citep{Planck2016}. This model is widely accepted as the standard framework for understanding the Universe. However, the precision era of cosmology has unveiled new challenges, such as $H_0$ tension between the CMB \citep{Planck2020} and the Supernova H0 for the Equation of State (SH0ES) results \citep{Riess2019} based on the cosmic distance ladder. Moreover, there is a $S_8$ tension between CMB \citep{Planck2020} and weak lensing measurement \citep{ KiDs2021, DES2022}. It is important to note that CMB measurements of cosmological parameters, including $H_0$, rely on the assumption of the $\Lambda$CDM model based on GR. This implies that there may be some unidentified systematic effects at play, which could be influencing the observed data or the $\Lambda$CDM model may not be fully adequate to describe the dynamics of our Universe, hinting at the possibility of new physics. Consequently, it is crucial to verify the validity of GR without referring to specific cosmological models. Significant strides have been made in testing GR within our solar system, such as the Cassini mission, which provided a highly accurate measurement of the parametrized post-Newtonian parameter (PPN) $\gamma_{\rm PPN}$ \citep{Verma2014} with a precision of approximately $10^{-5}$. However, theoretical ideas regarding the screening effect or chameleon mechanism, which could operate in some extensions of the GR, diminish the role of solar system tests. Therefore, it is essential to investigate potential deviations from GR in the non-linear regime, particularly on kiloparsec (kpc) scales. For example, observed tomographic weak lensing and redshift-space distortions indicated that there are roughly $5\%$ discrepancies in the Newtonian and curvature potentials when compared to GR predictions \citep{Simpson2013}.

The test of GR using strong gravitational lensing (SGL) is achieved through the measurement of two masses, namely the projected lens mass inside the Einstein radius and the kinematic mass from the central velocity dispersions of the lensing galaxy. For testing GR this way, please refer to \citet{Cao2017PPN,Wei2022}. The modification of basic GR theory can generally be realized by introducing the  $\gamma_{\rm PPN}$ into the standard Schwarzschild metric \citep{Thorne1971}. Subsequently, \citet{Yang2020} introduced an innovative methodology that combines time-delay measurements with the stellar kinematics of the lens to simultaneously constrain $H_0$ and $\gamma_{\rm PPN}$ from strong lensing observables. The results indicated no significant deviation from the GR with the obtained value of $\gamma_{\rm PPN}=0.87^{+0.19}_{-0.17}$. However, it is crucial to note that their work was cosmological model-dependent by assuming the $\Lambda$CDM model. Such tests should be conducted without relying on any specific cosmological model to minimize the influence of parametric or model-based assumptions. More recently, \citet{Liu2024} refined their approach by applying the Gaussian Process (GP) regression method to reconstruct distances based on the Pantheon dataset of SN Ia. This change was intended to provide a more robust and model-independent estimation of the distance, thereby enhancing the reliability of the results obtained from gravitational lensing studies. There is a caveat, however, that SN Ia directly provided luminosity distance $D_{\rm L}$ and in order to use them in the context of time delay distances from SGL data, it is necessary to convert them to the angular diameter distance $D_{\rm A}$ using the Etherington reciprocity formula \citep{Wang2025}. Even though well-grounded theoretically, such an approach may introduce some bias to the results. Therefore, identifying a sample that can directly reconstruct $D_{\rm A}$ could enhance the methodology proposed by previous studies. In this paper, we reconstruct the angular BAO scale $\theta$ using Gaussian Process Regression (GPR) based on the latest baryon acoustic oscillation (BAO) measurements, provided by the Dark Energy Spectroscopic Instrument Data Release 2 (DESI DR2) data. 

The structure of this paper is as follows. In Section \ref{sec:2}, we elaborate on the theoretical methodology of employing the SGL test for $\gamma_{\rm PPN}$. Section \ref{sec:3} introduces the GPR, which we apply in conjunction with BAO data sets from DESI DR2, for the reconstruction of the angular BAO scale $\theta$. The main conclusions and discussions are presented in Section \ref{sec:4}.

\section{Methodology}\label{sec:2}

The idea of using strong lensing systems to explore parametrized post-Newtonian corrections to GR is as follows. Assuming that the lensing system, i.e., the source, the lens, and the observer, are located in a homogeneous and isotropic universe, we treat the lens as a weak perturbation of the FLRW metric 
\begin{equation}
    {\rm d}s^2=-\left(1+\frac{2\Phi}{c^2}\right)c^2{\rm d}t^2+a(t)^2\left(1-\frac{2\Psi}{c^2}\right){\rm d}\bm{x}^2,
    \label{eq:1}
\end{equation}
where $c$ denotes the speed of light and $a(t)$ represents the scale factor. The metric is primarily influenced by the Newtonian potential $\Phi$ and the spatial curvature potential $\Psi$. In the weak-field approximation, GR posits $\Phi=\Psi$, and any divergence from this equality is quantified by the ratio $\gamma_{\rm{PPN}}=\Psi/\Phi$, which indicates the space-curvature per unit rest mass. A value of $\gamma_{\rm{PPN}}=1$ signifies the recovery of the GR. The deflection of light in the gravitational field of massive galaxies, a phenomenon anticipated by GR, makes SGL systems ideal for examining the distinction between the $\Phi$ and $\Psi$ on kpc scales. The Newtonian potential $\Phi$ is responsible for the motions of non-relativistic matter, such as baryonic and dark matter. Hence, the stellar dynamics of the lensing galaxy are sensitive solely to $\Phi$, which can be inferred from spectroscopic observations of central velocity dispersions. The lensing observables, including measurements of the time delay of multiple images in SGL systems, are related to both $\Psi$ and $\Phi$. Consequently, the Weyl potential,  $\Phi_+=\frac{\Phi+\Psi}{2}=(\frac{1+\gamma_{\rm {PPN}}}{2})\Phi$, is introduced to encapsulate the total potential, which can influence multiple-image measurements of SGL systems.

In the framework of GR, the time delay $\Delta t_{ij}$ between two images of the source located at angular position $\bm{\beta}$ relative to the center of the lens, seen at the angular positions $\bm{\theta}_i, \bm{\theta}_j$ is given by
\begin{equation}
    \begin{aligned}
    \Delta t_{ij}&=\frac{D_{\Delta t}}{c} \left[\frac{(\bm{\theta}_i - \bm{\beta})^2}{2}-\phi(\bm{\theta}_i)-\frac{(\bm{\theta}_j -\bm{\beta})^2}{2}+\phi(\bm{\theta}_j)\right]\\
    &=\frac{D_{\Delta t}}{c}\Delta\psi_{ij}(\xi_{\rm{lens}}),
    \end{aligned}
    \label{eq:2}
\end{equation}
The expression in the brackets is the difference of Fermat potentials $\psi(\bm{\theta}, \bm{\beta}) = (\bm{\theta} - \bm{\beta})^2/2-\phi(\bm{\theta})$. The effective lens potential $\phi(\bm{\theta})$, which is the integral of the Weyl potential along the line-of-sight, depends on the mass density distribution of the lens \citep{Meneghetti2021}. The Fermat potential difference $\Delta\psi_{ij}(\xi_{\rm{lens}})$ can be reconstructed by high-resolution lensing imaging from space telescopes. For this purpose, the Fermat potential is modeled using some function that contains adjustable parameters $\xi_{\rm{lens}}$ that measure, e.g., the depth of the potential, its core radius, its ellipticity, and its radial variation. The factor $D_{\Delta t}$ in Eq.(\ref{eq:2}) is called the time-delay distance and is defined as $ D_{\Delta t} \equiv (1+z_{\rm l})\frac{D_{\rm l}D_{\rm s}}{D_{\rm ls}}$, where $z_{\rm l}$ is the redshift of the lens, $D_{\rm l}$ is the angular diameter distance to the lens, $D_{\rm s}$ is the angular diameter distance to the source, and $D_{\rm ls}$ is the angular diameter distance between the lens and the source. 

In the framework of modified gravity, when the Weyl potential is not equivalent to the Newtonian one and $\gamma_{\rm{PPN}} \neq 1$, the actually inferred lens model parameters $\xi'_{\rm{lens}}$ in the Fermat potential are different from the analogous parameters $\xi_{\rm{lens}}$ that would be derived under GR. Hence, when MG influences the lensing image reconstruction, the inferred time-delay distance is 
\begin{equation}
    D_{\Delta t} =\frac{c\Delta t_{ij}}{\Delta \psi_{ij}(\xi'_{\rm{lens}})}.
    \label{eq:3}
\end{equation}
One may benefit from including stellar kinematics of the lens since it depends only on the Newtonian potential $\Phi$ and is independent of $\gamma_{\rm PPN}$. Following the approach of the H0LiCOW collaboration to incorporate stellar kinematics, we introduce another observable $\sigma_v$, which represents the line of sight (LOS) projected stellar velocity dispersion of the lensing galaxy. This observable, together with measurements of multiple image positions, allows us to obtain another distance ratio 
\begin{equation}
    \frac{D_{\rm s}}{D_{\rm ls}}=\frac{\sigma_v^2}{c^2J(\xi_{\rm lens},\xi_{\rm light},\beta_{\rm ani})},
    \label{eq:4}
\end{equation}
where $\xi_{\rm lens}$ denote lens model parameters, $\xi_{\rm light}$ denote parameters related to luminosity distribution of the lens and $\beta_{\rm ani}$ represents the stellar anisotropy parameter. The appearance of these additional parameters is motivated by the fact that the anisotropic Jeans equation can model velocity dispersion $\sigma_v$, and the $J$ function encompasses all of the model components calculated from the lensed images and luminosity-weighted projected velocity dispersion \citep{koopmans2006gravitational,2015ApJ...806..185C}. When $\gamma_{\rm PPN}$ is introduced, it will modify the lens model parameters $\xi'_{\rm lens}$ inferred from time delays. Now this ratio is rescaled to
\begin{equation}
    \frac{2}{1+\gamma_{\rm PPN}}\frac{D_{\rm s}}{D_{\rm ls}}=\frac{\sigma_v^2}{c^2J(\xi'_{\rm lens},\xi_{\rm light},\beta_{\rm ani})},
    \label{eq:ratio_scale}
\end{equation}
Through the combination of Eqs.~(\ref{eq:3}) and (\ref{eq:ratio_scale}), the angular diameter distance to the lens can be expressed as
\begin{equation}
    D'_{\rm l} = \frac{1}{1+z_{\rm l}}\frac{c\Delta t_{ij}}{\Delta \psi_{ij}(\xi'_{\rm lens})}\frac{c^2J(\xi'_{\rm lens},\xi_{\rm light},\beta_{\rm ani})}{\sigma^2},
    \label{eq:5}
\end{equation}
which relates to its counterpart within the GR framework as
\begin{equation}
    D_{\rm l}'=\frac{1+\gamma_{\rm PPN}}{2}D_{\rm l}.
    \label{eq:add_gamma}
\end{equation}
If one is able to confront $D_{\rm l}'$ and $D_{\Delta t}$ with model-independent assessment of the angular diameter distance $D_{\rm l}$ and the time delay distance $D_{\Delta t}$, we could derive the direct measurement of the PPN parameter.

\section{Data and analysis}\label{sec:3}

The H0LiCOW collaboration has conducted an analysis of six lens systems spanning a broad range of lens and source redshifts \citep{H0LiCOW2020}. Four lens systems, i.e., RXJ1131-1231, PG1115+080, B1608+656, and J1206+4332, provided the measurements for both the $D_{\Delta t}$ and $D_{\rm l}$. It is worth noting that only the earliest system, B1608+656, has independent parametric skewed log-normal posterior distribution functions (PDFs) of $D_{\Delta t}$ \citep{Suyu2010} and $D_{\rm l}$ \citep{Jee2019} on the H0LiCOW website \footnote{\url{ http://www.h0licow.org}}. The PDFs of the other three lenses were released in the form of MCMC chains \citep{Suyu2014, Birrer2019, Chen2019}, which preserved the correlations between $D_{\Delta t}$ and $D_{\rm l}$. Although it is not easy to measure the angular diameter distances directly, the $D_{\rm A}(z)$ function can be reconstructed using a certain type of standard rulers. Assuming a spatially flat universe, the angular diameter distance between the lens and the source $D_{\rm ls}$ can be computed as $D_{\rm ls} = D_{\rm A}(z_{\rm s})-(1+z_{\rm l})/(1+z_{\rm s})D_{\rm A}(z_{\rm l})$ in terms of distances to the lens and to the source, which could be assessed from the reconstructed $D_{\rm A}(z)$ function \citep{2019PhRvD.100b3530Q,2019NatSR...911608C}.

The clustering of matter induced by BAO provides a standard length scale in cosmology. This scale, approximately $150~\mathrm{Mpc}$ at the present epoch, is determined from astronomical surveys of the large-scale structure, allowing constraints on cosmological parameters -- particularly the baryonic matter density -- and offering insights into the nature of dark energy driving the accelerated expansion of the Universe. The accuracy of this standard ruler is critical for cosmological studies. The angular BAO scale can be expressed as
\begin{equation}
\theta(z)=\frac{r_{\rm d}}{D_{\rm M}(z)},
\label{eq:6}
\end{equation}
where $r_{\rm d}$ denotes the sound horizon at redshift $z$ and ${D_{\rm M}(z)}= (1+z){D_{\rm A}(z)}$. Now, the determination of $D'_{\rm l}$ depends on both the post-Newtonian parameter $\gamma_{\rm PPN}$ and the sound horizon at drag epoch $r_{\rm d}$. More specifically, the combination of Eq.~(\ref{eq:add_gamma}) and (\ref{eq:6}) generates the modified angular diameter distance as
\begin{equation}
    D'_{\rm l}=\frac{1+\gamma_{\rm PPN}}{2}\frac{r_{\rm d}}{(1+z)\theta(z)}.
    \label{eq:add}
\end{equation}

As part of the BAO analysis, the most recent results from the DESI DR2 \citep{DESI2025} are first incorporated. The measurements of transverse comoving distance $D_{\rm M}$ are derived from Luminous Red Galaxies (LRGs), Emission Line Galaxies (ELGs), quasars (QSOs), and the Lyman-$\alpha$ forest (Ly$\alpha$) over the redshift range $0.4 < z < 4.2$, with effective redshifts varying from 0.51 to 2.33. The LRG and ELG tracers are divided into four non-overlapping redshift bins, so that their $D_{\rm M}$ measurements are analyzed separately. Given that the bright galaxy sample (BGS) at lower redshift cannot provide tight constraints on $D_{\rm M}$ owing to its low signal-to-noise ratio in quadrupole fitting, we supplement our sample with additional measurements from other surveys to ensure that the data points used for reconstruction cover the full redshift range spanned by all strong gravitational lensing (SGL) systems. Fortunately, the Sloan Digital Sky Survey (SDSS) satisfies this requirement: its Data Release 16 (DR16) provides the final data release for the main cosmological program of the Extended Baryon Oscillation Spectroscopic Survey (eBOSS), which also incorporates data from its predecessor, the Baryon Oscillation Spectroscopic Survey (BOSS) \citep{SDSSDR162020}. From SDSS DR16, we include four additional anisotropic BAO measurements in our dataset, from which $D_{\rm M}$ values are extracted for LRGs \citep{SDSSLRG}, ELGs \citep{SDSSELG}, quasars \citep{SDSSQSO}, and Ly$\alpha$ forests \citep{SDSSlya}, respectively. Furthermore, the LOWZ sample from the earlier SDSS DR12 selects galaxies in the redshift range $0.15 \le z \le 0.43$ and was not updated in DR16, yielding an additional data point at at $z_{\rm eff} = 0.32$ \citep{SDSS0.32}. Finally, two projected transverse BAO $\theta$ measurements from SDSS and DES are also incorporated, in order to extend the redshift coverage and enhance the robustness of GPR reconstruction \citep{SDSS0.11, DES:2024cme}. In summary, the dataset used for reconstruction comprises eleven anisotropic and two transverse BAO points. All data points from the same survey are drawn from non-overlapping redshift bins, ensuring that all BAO measurements can be treated independently in the subsequent reconstruction without the need to account for any covariance.

\begin{figure}[!t]
{\includegraphics[width=1.1\linewidth,height=\linewidth]{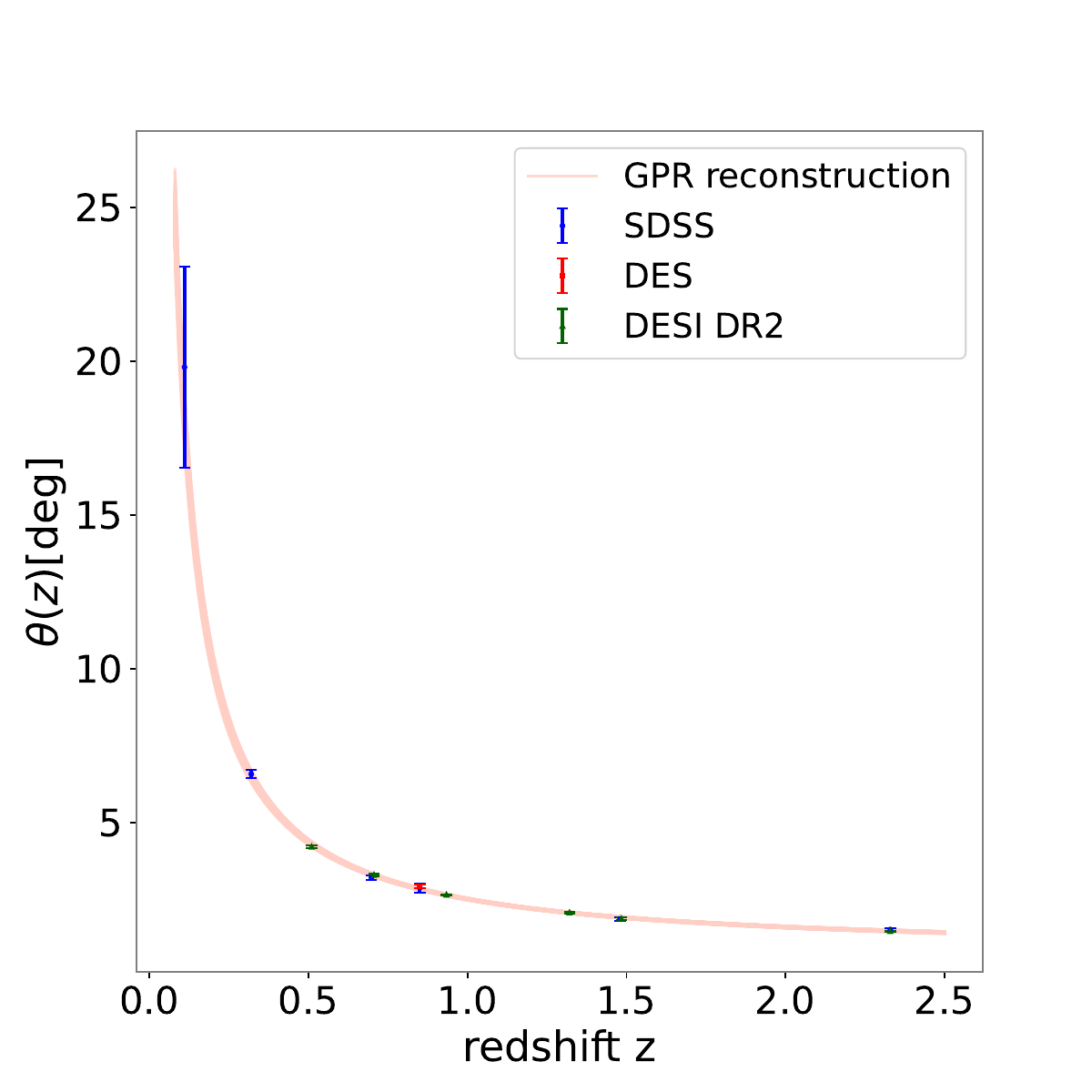}}
\caption{The measurements of BAO angular scale from DESI DR2 and other galaxy surveys. The light pink lines are the GPR reconstructions consisting of 1000 curves.}\label{fig:BAO}
\end{figure}

In order to derive the measurement of $\gamma_{\rm PPN}$ from strong lensing systems, distance information at the lens and source redshifts is required. Given the limited overlap between BAO and lensing datasets, we employ GPR, a cosmology-independent reconstruction technique, to recover the angular BAO scale \citep{Holsclaw1, Keeley0, ShafKimLind}. The reconstruction is performed using posterior samples from combined 2D and 3D BAO measurements, generated with the \texttt{GPHist}\footnote{\url{https://github.com/dkirkby/gphist}} code \citep{GPHist}. GPR is a fully data-driven method that operates in an infinite-dimensional function space and mitigates overfitting \citep{Keeley0}. It models the reconstructed function $\gamma(z)$ as a distribution of functions defined by a covariance structure, for which we adopt the commonly used squared-exponential kernel
\begin{equation}
    \langle \gamma(z_1)\gamma(z_2) \rangle = \sigma_f^2 \, \exp\{-[s(z_1)-s(z_2)]^2/(2\ell^2)\},
\end{equation}
where $\sigma_f$ and $\ell$ are hyperparameters that would be marginalized over. The function $\gamma(z)$ is drawn from the distribution defined by the covariance, and we set $\gamma(z) = \ln[\theta_{\mathrm{BAO}}(z)]$ to generate additional angular scale measurements from the BAO dataset. A total of 1000 reconstructed curves are obtained, as shown in Fig.~\ref{fig:BAO}, which capture the angular scale–redshift relation of the BAO data well. The BAO redshift coverage encompasses that of the H0LiCOW lensing systems, eliminating the need for extrapolation.

Based on Eqs.~(\ref{eq:3}) and (\ref{eq:add}), all we need to do is to establish the relation between reconstruction $\theta(z)$ and $D_{\Delta t}$ as well as $D'_{\rm l}$. It should be stressed that only the lens distance ($D'_{\rm l}$) carries the information of the PPN parameter, while the time-delay distance ($D_{\Delta t}$) is sensitive to $r_{\rm d}$. Thus, the combination of $D_{\Delta t}$ and $D_{\rm l}$ directly provides an alternative approach for the simultaneous measurement of $r_{\rm d}$ and $\gamma_{\rm PPN}$. First of all, we calculate 1000 sets of scaled $r_{\rm d}/D_{\rm l}$ and $r_{\rm d}/D_{\Delta t}$ based on GPR reconstructions for each lensing system. Then, utilizing the lattice point method, the 2D-PDFs, after marginalizing over 1000 realizations, in the parameter space of $r_{\rm d}$ and $\gamma_{\rm PPN}$ for each lens are evaluated using the constraints on $D_{\Delta t}$ and $D_{\rm l}$ from the H0LiCOW. The joint probability distribution for the total four lenses is constructed by multiplying the probabilities for each of the 1000 individual GPR reconstruction curves across the systems. Such an approach could preserve and propagate the intrinsic correlations within the data points of each GPR curve, thereby ensuring they are fully accounted for in the final probability distribution \citep{Liu2025}. Finally, the posterior probability distributions of $r_{\rm d}$ and $\gamma_{\rm PPN}$ can be derived simultaneously.

\section{Results and Discussion}\label{sec:4}

\begin{figure}[!t]
  \centering
    \includegraphics[width=0.5\textwidth]{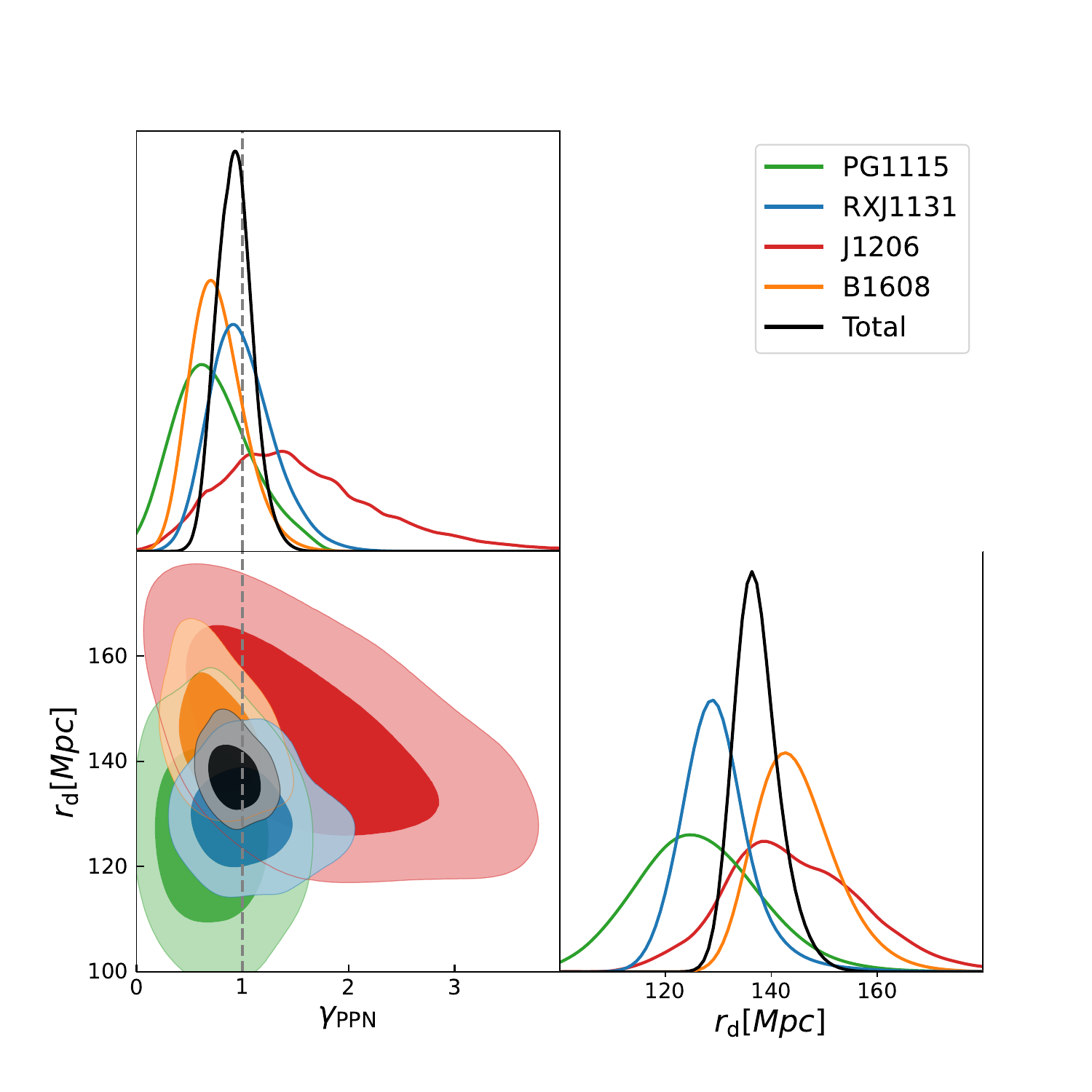}
  \caption{The 1D probability distributions and 2D contours with
1$\sigma$ and 2$\sigma$ confidence levels, for the post-Newtonian parameter ($\gamma_{\rm{PPN}}$) and the sound horizon ($r_{\rm d}$). The dashed line represents the prediction of GR ($\gamma_{\rm{PPN}}=1$).}
  \label{fig:gard}
\end{figure}

Working on the four posteriors of $D_{\Delta t}$ and $D_{\rm l}$ released by H0LiCOW, along with the reconstructed angular diameter distances from BAO dataset, we obtain the final constraints on the PPN parameter $\gamma_{\rm PPN}$ and sound horizon scale $r_{\rm d}$. The 1D probability distributions and 2D contours with 1$\sigma$ and 2$\sigma$ confidence levels are shown in Fig.~\ref{fig:gard}. The degeneracy between $D_{\Delta t}$ and $D_{\rm l}$ varies significantly across different lensing systems. Based on the individual measurements from four lensing systems listed in Table~\ref {rdandga}, we directly estimate the post-Newtonian parameter $\gamma_{\rm PPN}=0.93^{+0.16}_{-0.17}$ and the sound horizon scale $r_{\rm d}=136.36^{+5.14}_{-3.20}~{\rm Mpc}$. This is the first simultaneous measurement of $\gamma_{\rm PPN}$ and $r_{\rm d}$ without any assumptions about the contents of the universe or the theory of gravity.

Although the central value of $\gamma_{\rm PPN}$ is slightly different from 1, our results still support GR within $1\sigma$. Actually, the past two decades have seen great progress in measuring the post-Newtonian parameter with a variety of cosmological probes. For instance, by comparing the masses of 15 elliptical lensing galaxies from the Sloan Lens ACS Survey (SLACS), \citet{Bolton2006} tested the weak-field metric on kiloparsec scales and found a constraint of $\gamma_{\rm PPN}=0.98\pm0.07$. 
Such methodology was extended to a mass-selected sample of galaxy-scale lenses from the SLACS, BELLS, LSD, and SL2S surveys \citep{Cao2015}. Based on a well-motivated fiducial set of lens-galaxy parameters, the measurement of the post-Newtonian parameter changes to $\gamma_{\rm PPN}=0.995^{+0.037}_{-0.047}$ \citep{Cao2017PPN}. More recently, by combining the updated observations of SGL and SN Ia, \citet{Wei2022} directly estimated the post-Newtonian parameter $\gamma_{\rm PPN}=1.11^{+0.11}_{-0.09}$ and the cosmic curvature $\Omega_k=0.48^{+1.09}_{-0.71}$. The anisotropies in the Cosmic Microwave Background can also be used to perform the tests of GR. \citet{Thomas2024} investigated the $\gamma_{\rm PPN}$ between last scattering and the present day, determining their time-averages of $\bar{\gamma}_{\rm PPN}=0.90^{+0.07}_{-0.08}$ over cosmological history. Our joint result agrees with the above results, showing great consistency between different observations. On the other hand, our constraint on $r_{\rm d}$ achieves a precision of $\sim 2.1\%$, and shows a tension with the Planck 2018 result ($r_{\rm d} = 147.09 \pm 0.26 \,\mathrm{Mpc}$) within $2.7\sigma$. Analogously, based on the combined SGL+SN Ia+BAO data sets, \citet{Bernalrd136} also derived a low value of $r_{\rm d} = 136.8\pm{4.0}~{\rm Mpc}$, using the spline interpolation method to reconstruct the cosmic expansion history. \citet{Liu2025rd} obtained a model-independent result of $r_{\rm d}=139.5\pm5.2~{\rm Mpc}$ using the observations of DESI DR1 and time-delay lensed quasars from H0LiCOW. Our findings are well consistent with the previous analysis based on other low-redshift observations, which is also in tension with the high-redshift constraints from Planck. This could be another indication of the well-known Hubble tension, considering the strong correlation between $r_{\rm d}$ and $H_0$.
\begin{table}[!t]
\renewcommand\arraystretch{1.3}
\caption{Constraints on the post-Newtonian parameter ($\gamma_{\rm PPN}$) and sound horizon ($r_{\rm d}$) from four H0LiCOW lenses.}
\begin{center}
\begin{tabular}{l| c c }
\hline
\hline
Lens   &$\gamma_{\rm PPN}$  & $r_{\rm d} ({\rm Mpc})$
\\
\hline
All   & $0.93^{+0.16}_{-0.17}$  & $136.36^{+5.14}_{-3.20}$\\
\hline
PG1115  & $0.61^{+0.48}_{-0.24}$ & $124.56^{+13.49}_{-9.61}$  \\
\hline
RXJ1131  & $0.91^{+0.39}_{-0.21}$ & $129.08^{+6.64}_{-5.62}$  \\
\hline
J1206  & $1.37^{+0.90}_{-0.54}$ & $139.08^{+17.98}_{-7.23}$  \\
\hline
B1608   & $0.70^{+0.31}_{-0.18}$ & $142.72^{+9.71}_{-5.27}$ \\
\hline
\hline
\end{tabular}
\end{center}
\label{rdandga}
\end{table}

\begin{table}
\renewcommand\arraystretch{1.3}
\caption{Constraints on $\gamma_{\rm PPN}$ using the distance ratio method from four H0LiCOW lenses.}
\begin{center}
\setlength{\tabcolsep}{1.5mm}{
\begin{tabular}{l| c c c c c}
\hline
\hline
Lens  & All  &PG1115 & RXJ1131 &J1206 & B1608 \\
\hline
$\gamma_{\rm PPN}$ & $0.89^{+0.19}_{-0.15}$ & $0.73^{+0.53}_{-0.23}$  & $0.96^{+0.45}_{-0.18}$ & $1.45^{+1.34}_{-0.33}$ & $0.70^{+0.31}_{-0.18}$ \\
\hline
\hline
\end{tabular}}
\end{center}
\label{ratio}
\end{table}

\begin{figure}[!t]
\centering
\includegraphics[width=8.5cm,height=7.8cm]{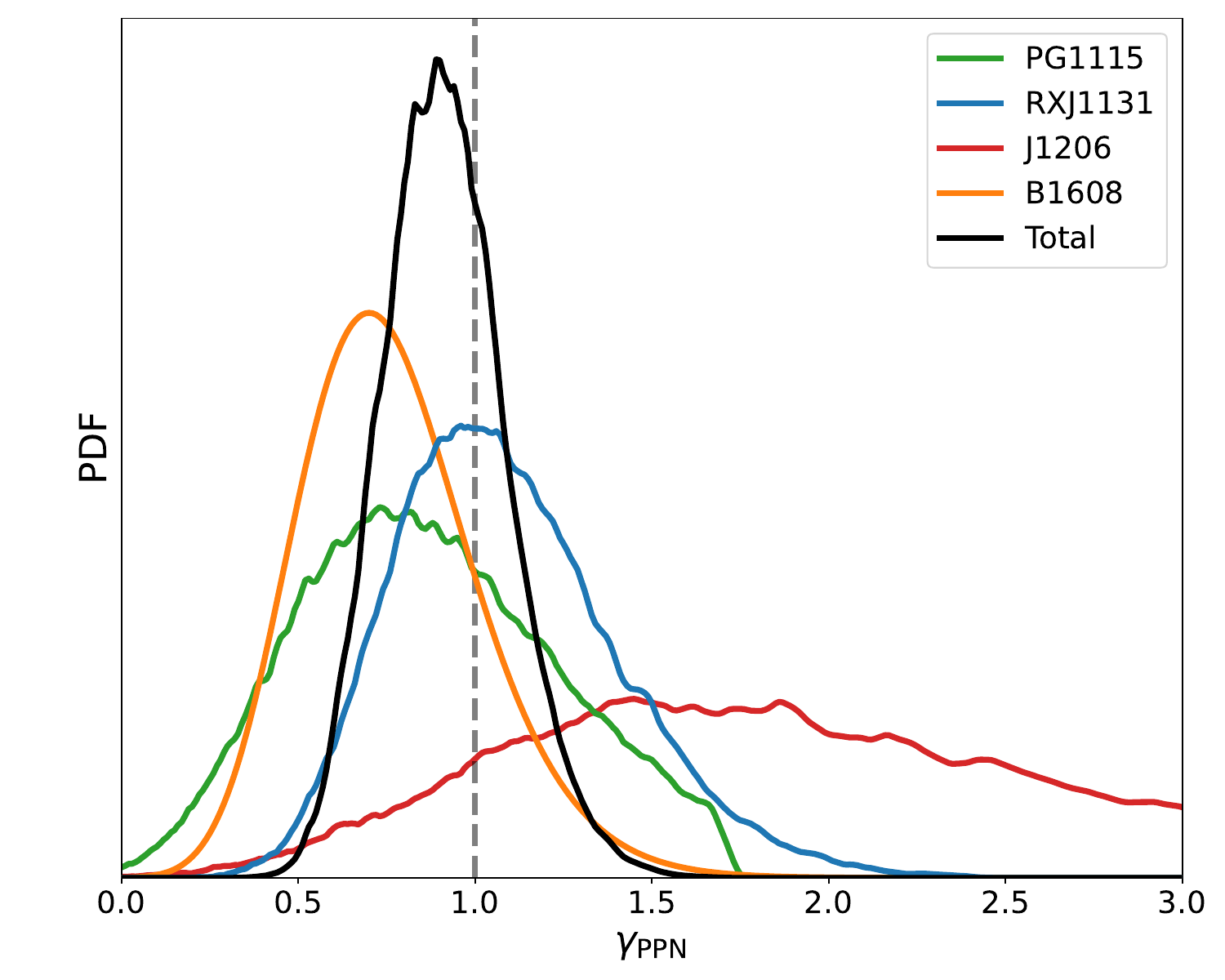}
\caption{The one-dimensional posterior distributions of $\gamma_{\rm{PPN}}$ using the distance ratio method. The dashed line is $\gamma_{\rm{PPN}}=1$ predicted by GR.}
\label{ratioga}
\end{figure}

In the analysis above, the value of $D_{\rm l}$ is determined by both $\gamma_{\rm PPN}$ and $r_{\rm d}$, which leads to an unavoidable degeneracy between the two parameters. As a complementary approach, using the distance ratio 
\begin{equation}
 \frac{D_{\Delta t}}{D'_{\rm l}} = \frac{2}{1+\gamma_{\rm PPN}} \frac{D_{\rm s}}{D_{\rm ls}}   
\end{equation}
could effectively avoid the bias induced by $r_{\rm d}$. Here the $D_{\rm s}/D_{\rm ls}$ ratio is assessed from of reconstructed angular scale $\theta(z)$ (Fig.~\ref{fig:BAO}), i.e., $D_{\rm s}/D_{\rm ls}=\theta(z_{\rm l})/[\theta(z_{\rm l})-\theta(z_{\rm s})]$. Such methodology could be traced back to the papers of \citet{Liu2024}, which discussed the possibility of deriving the angular-diameter-distance ratio $D_{\rm ls}/D_{\rm s}$ from reliable knowledge of the lensing system (i.e., measurements of Einstein radius and central velocity dispersion). In the framework of such methodology, the joint constraint on the post-Newtonian parameter is $\gamma_{\rm PPN}=0.89^{+0.19}_{-0.15}$, which also agrees well with the prediction of GR (see Fig.~\ref{ratioga} and Table~\ref{ratio}). However, one should point out that, since the $D_{\Delta t}$ and $D_{\rm l}$ measurements given by H0LiCOW are correlated except for B1608+656, the correlation between the two distance measurements would be one source of systematics. Moreover, since the reconstruction methods do not significantly improve the precision of distance reconstruction from BAO \citep{Zheng2025}, we are still looking forward to new available time-delay gravitational lenses to derive more stringent constraints on $\gamma_{\rm PPN}$. With 160000 galaxy-scale strong lensing systems detected by the China Space Station Telescope (CSST) \citep{2024MNRAS.533.1960C}, one could also expect direct tests of GR under screening effect \citep{2022ApJ...941...16L}.

\section*{Acknowledgements} 

This work is supported by Beijing Natural Science Foundation No. 1242021; the National Natural Science Foundation of China (Nos. 12021003, 12203009, 12433001); the Strategic Priority Research Program of the Chinese Academy of Sciences, Grant No. XDB23000000; the Interdiscipline Research Funds of Beijing Normal University; and the Fundamental Research Funds for the Central Universities.

\bibliographystyle{spphys}
\bibliography{reference.bib}
\end{document}